\documentclass[fleqn,a4paper,12pt]{article}
\usepackage{amssymb}
\usepackage{amsmath}
\usepackage[dvips]{graphicx}
\usepackage{epsfig}

\newcommand{\bab}{\end{gather}}
\newcommand{\ri}{{\mathrm i}}
\newcommand{\p}{\partial}
\newcommand{\bea}{\begin{array}}
\newcommand{\eea}{\end{array}}
\newcommand{\beg}{\begin{gather}}

\catcode`@=11 \long
\def\@caption#1[#2]#3{\par\addcontentsline{\csname
ext@#1\endcsname}{#1} {\protect\numberline{\csname
the#1\endcsname}{\ignorespaces #2}} \begingroup \small
\@parboxrestore \@makecaption{\csname fnum@#1\endcsname}
{\ignorespaces #3}\par \endgroup} \catcode`@=12

\newcommand{\Q}{\mathbb{Q}}

\newcommand{\la}{\label}

\catcode`@=11 \long
\def\@caption#1[#2]#3{\par\addcontentsline{\csname
ext@#1\endcsname}{#1} {\protect\numberline{\csname
the#1\endcsname}{\ignorespaces #2}} \begingroup \small
\@parboxrestore \@makecaption{\csname fnum@#1\endcsname}
{\ignorespaces #3}\par \endgroup} \catcode`@=12

\begin{document}

\allowdisplaybreaks
 \begin{titlepage} \vskip 2cm

\begin{center} {\Large\bf Superintegrable quantum mechanical systems with position dependent
masses invariant with respect to three parametric Lie groups}

 \vskip 3cm {\bf {A. G. Nikitin }\footnote{E-mail:
{\tt nikitin@imath.kiev.ua} }
\vskip 5pt {\sl Institute of Mathematics, National Academy of
Sciences of Ukraine,\\ 3 Tereshchenkivs'ka Street, Kyiv-4, Ukraine,
01024, and\\Universit\'a del Piemonte Orientale,\\
Dipartimento di Scienze e Innovazione Tecnologica,\\
viale T. Michel 11, 15121 Alessandria, Italy}}\end{center}
\vskip .5cm \rm
\begin{abstract} { Quantum mechanical
systems with position dependent masses (PDM) admitting four and more dimensional symmetry
algebras  are classified. Namely,
all PDM  systems are specified which, in addition to their invariance w.r.t. a three
parametric Lie group, admit at least one second order integral of motion. The presented classification
is partially extended to the more generic systems which admit one or two parametric Lie groups. }
\end{abstract}
\end{titlepage}
\section{Introduction\label{int}}

The title of the present paper is a bit conventional. The results presented there are more generic than
it declares. In addition to the PDM systems admitting three parametric Lie groups and second order
integrals of motion we give the classification of the systems invariant with respect to  selected two-
and one parametric groups.

Let us start with short historical comments related to symmetries of quantum mechanical systems
with constant masses.

 Symmetry is one of the most fundamental concept of theoretical and mathematical physics, especially  of quantum mechanics. The fundamentals of the science of continuous
 symmetries were created long time ago by the great Norwegian mathematician Sophus Lie.  In
 particular de facto he  discovered all such symmetries admitted by the fundamental  equation of
 quantum mechanics. More exactly, Lie found the maximal continuous invariance group of the heat
 equation, which in the main coincides with the symmetry group of the free Schr\"odinger equation.

A systematic search for Lie symmetries of Schr\"odinger equation started  in papers \cite{Hag, Nied,
And} and \cite{Boy} where the maximal invariance groups of this equation with arbitrary scalar
potential were presented. For symmetries of this equation with scalar and vector potentials and
corrected results of classical paper \cite{Boy}  see  papers \cite{N22, N32}, Lie symmetries of
Schr\"odinger equation with matrix potentials are classified in \cite{N42, N52}.

The more general  symmetries, namely, the second order symmetry operators for 2$d$ and 3$d$
Schr\"odinger equation have been classified in \cite{wint1}, \cite{BM} and \cite{ev1}, \cite{ev2}
correspondingly. The extended (in particular, second order) symmetries are requested
for  description of systems admitting solutions in separated
variables \cite{Miller} . Just such symmetries characterize integrable and superintegrable systems
\cite{wint01}. Let us mention also the nice conjecture of  Ian Marquette and Pavel Winternitz
\cite{Mar}
which can give a surprising connection of higher order
superintegrability in the quantum case with soliton theory of infinite-dimensional integrable
nonlinear systems.

An important research field is formed by
superintegrable systems with spin whose systematic investigation
was started with paper \cite{w6, w7, w8} where the systems with spin-orbit interaction were classified.
Superintegrable  systems with Pauli type interactions were studied
in \cite{N5, N1} and  \cite{N2}.

Let us note that the first example of a superintegrable system with spin 1/2 was presented earlier in
paper \cite{Pron1}. Superintegrable systems with arbitrary spin were discussed in \cite{Pron2},
\cite{N1}, \cite{N6} and \cite{N5}, the relativistic systems were elaborated in \cite{N2} and \cite{N77}.

The modern trend is to study  the  superintegrable systems admitting integrals of motion
of the third and even arbitrary orders \cite{Mar, wintL}, see also  \cite{AGN1} where the determining
equations for such symmetries were deduced, and \cite{N7} where symmetry operators of arbitrary
order for the free Schr\"odinger equation had been enumerated.

Thus the amazing world of symmetries of Schr\"odinger equation is an important and interesting
research field which attracts the attention of numerous investigators. The same is true for the
Schr\"odinger equation with position dependent mass whose symmetries are studied much less. The
latter equation is requested in many branches of modern theoretical physics, whose list can be found,
e.g., in   \cite{Rosas, NZ}.

Symmetries of various PDM Schr\"odinger equations with respect to the continuous groups have been
classified in papers \cite{NZ} -\cite{AGN}. More exactly, the symmetries of the stationary equation are
presented in \cite{NZ} while the time dependent equations with two and three spatial variables are
studied in \cite{NZ2} and \cite{AGN} correspondingly.

The situation with the higher symmetries of the PDM quantum mechanical systems is much  more
complicated. There is a lot of paper devoted to symmetries of particular equations or of the restricted
classes of such equations, see, e.g., \cite{Bala2, Rag1, Ra, Ca, Axel, 154,Vol}.   However, the completed
classification of 3d superintegrable systems with PDM is still missing.

On the other hand the 2d classical systems with position dependent mass which admit second order
integrals of motion  are known and well studied \cite{Koen,Kal, Kal2, kama, kaka} and there is a   correspondence between classical
and quantum
 superintegrable systems \cite{Kal3}.

 The main stream  in studying of superintegrable systems with PDM is the investigation of classical
 Hamiltonian systems. And  there are effective tools for such business created in classical works of Bernard, St\"ackel,
 Koenigs and Perlick.

 Surely there exist the analogous  quantum mechanical systems which in
 principle can be obtained starting with the classical ones and applying the second quantization
 procedure. However, the mentioned procedure is not unique, and in general    it is possible to
 generate few inequivalent quantum systems which have the same classical limit. In addition, a part of
 symmetries and integrals  of motion of quantum mechanical systems can disappear in the classical limit $h\to 0$ \cite{Hit}.

 Thus it is desirable to  classify
 superintegrable quantum systems directly. However, to obtain the completed classification of such
 systems  is very and very difficult, and it is reasonable to solve this problem step by step, restricting
 ourselves to some well defined subclasses of such equations. And this is just the strategy which we
 will follow.

 In the present paper the complete classification of a special class of superintegrable PDM
 Schr\"odinger equations is presented. This class includes equations which admit three parametric
 symmetry groups. In addition, we will specify a certain subclass of such equations which admit the
 symmetry groups including two parameters.

\section{PDM Schr\"odinger equations}
We will search for superintegrable  stationary  Schr\"odinger equations with position dependent mass of
the following generic form:
\begin{gather}\la{se}
   H \psi=E \psi,
\end{gather}
where
\begin{gather}\la{H} H=p_af({\bf x})p_a+ V({\bf x}).\end{gather}
Here ${\bf x}=(x^1,x^2,x^3),$ $p_a=-i\p_a$, $V({\bf x})$ and $f({\bf x})=\frac1{2m({\bf x})}$ are
functions associated with the effective potential and inverse  PDM, and summation from 1 to 3 is
imposed over the repeating index $a$.

A more general  form of the PDM Hamiltonian is \cite{Roz}
\begin{gather}\la{A1}  H=\frac14(m^\alpha p_a m^{\beta}p_am^\gamma+
m^\gamma p_a m^{\beta}p_am^\alpha)+ \hat V\end{gather}
where  $\alpha, \beta$  and $\gamma$ are the so called
ambiguity parameters satisfying the condition $\alpha+\beta+\gamma=-1$. Physically, representation
(\ref{A1}) is more consistent
but mathematically it is completely equivalent to  (\ref{H}) \cite{AGN}.

In paper \cite{NZ} all equations (\ref{se}) admitting at least one first order integral of motion has been
classified. Such integrals of motion are nothing but generators of Lie groups which leave the related
equations invariant.  The list of such equations includes three representatives which accept three
parametrical invariance groups. The corresponding inverse masses $f$ and potentials $V$ are
presented in the following formulae:
\begin{gather}\la{f_V1}f=F(r), \quad V=V(r),\\\la{f_V2}f=F(x_3), \quad V=V(x_3),\\\la{f_V3}f=\tilde
r^2F(\varphi),\quad V=V(\varphi)\end{gather}
where $F(.)$ and $V(.)$ are arbitrary functions whose arguments are fixed in the brackets,
\begin{gather*} r=(x^2_1+x_2^2+x_3^2)^\frac12,\quad  \tilde r=(x^2_1+x_2^2)^\frac12,\quad
\varphi=\arctan\left(\frac{x_2}{x_1}\right).\end{gather*}

Equations (\ref{se}), (\ref{H}) whose arbitrary parameters are fixed by formulae (\ref{f_V1}), (\ref{f_V2})
and (\ref{f_V3}) admit the following integrals of motion
\begin{gather}\la{IM1}L_1=x_2p_3-x_3p_2, \quad L_2=x_3p_1-x_1p_3,\quad
L_2=x_3p_1-x_1p_3,\\\la{IM2}P_1=p_1,\quad P_2= p_2,\quad L_3\end{gather} and
\begin{gather}\la{IM3}P_3=p_3, \quad  D={ x_a}{ p_a}-\frac{3\ri}2,\quad K_3=x^2 p_3
-2x_3D\end{gather}
correspondingly, which form bases of Lie algebras so(3), e(2) and so(1,2) respectively. In other words,
equations  (\ref{se}), (\ref{H}), (\ref{f_V1})  and (\ref{se}), (\ref{H}),(\ref{f_V2}) are invariant w.r.t. the
rotation group SO(3) and Euclid group E(2) correspondingly while equations (\ref{se}), (\ref{H}),
(\ref{f_V1}) are invariant w.r.t. the three parametrical Lie group isomorphic to Lorentz group SO(1,2) in
(1+2) - dimensional space.

\section{Determining equations}
Let us search for second order integrals of motion for equation (\ref{se}), i.e., for second order
differential operators commuting with $H$. We will represent these integrals of motion in  the
following  form:
\begin{equation}\label{Q}
    Q=\p_a\mu^{ab}\p_b+\eta
\end{equation}
where $\mu^{ab}=\mu^{ba}$ and $\eta$ are unknown functions  of $\bf
x$ and summation from 1 to 3 is imposed over all repeating indices.

By definition, operators  $Q$ should commute with $H$:
\begin{equation}\label{HQ}[ H,Q]\equiv  H Q-Q H=0.\end{equation}
Evaluating the commutator and equating to zero the coefficients for the linearly independent differential operators $\p_a\p_b\p_c$ and $\p_a$ we come to the following determining equations:
\begin{gather}\la{de1}(f\mu^{ab}_c-\mu^{an}f_n\delta^{bc})+cycle(a,b,c)=0,
\\\la{mmmu4}\mu^{ab}V_b-f\eta_a=0\end{gather}
  where $\delta^{bc}$ is the Kronecker delta, $f_n=\frac{\p f}{\p x_n}, \ \mu^{an}_n=\frac{\p \mu^{an}}{\p x_n }$, etc., and summation is imposed over the repeating indices $n$, $n=1,2,3$.

 Equations (\ref{de1}) and (\ref{mmmu4})  present the necessary and sufficient conditions for commutativity of operators $H$  and $Q$.

 It is an element of common knowledge that a commutator of two second order differential operators is a linear combination of the third, second, first and an zero order operators. The beauty of the representations (\ref{H}) for $H$ and (\ref{Q}) for $Q$  is that we are not supposed to collect and nullify the coefficients for second and zero order differentials
 which do not appear in the commutator (\ref{HQ}).

Let us present also the traceless part and the trace of the tensorial equations (\ref{de1}):
\begin{gather}\la{mmmu0}5\left(\mu^{ab}_c+\mu^{ac}_b+ \mu^{bc}_a\right)=
\delta^{ab}\left(\mu^{nn}_c+2\mu^{cn}_n\right)+
\delta^{bc}\left(\mu^{nn}_a+2\mu^{an}_n\right)+\delta^{ac}
\left(\mu^{nn}_b+2\mu^{bn}_n\right),\\
\la{mmmu1}
 \left(\mu^{nn}_a+2\mu^{na}_n\right)f-
5\mu^{an}f_n=0.\end{gather}

Thus to classify Hamiltonians (\ref{H}) admitting second order
integrals of motion (\ref{Q}) we are supposed  to find
inequivalent solutions of rather complicated system
(\ref{mmmu0})--(\ref{mmmu4}).

The autonomous subsystem (\ref{mmmu0}) defines the conformal Killing tensor.
Its general solution is a linear combination of the following tensors
(see, e.g., \cite{Kil})
\begin{gather}\la{mmu1}\begin{split}&\mu^{ab}_0=\delta^{ab}g_0({\bf x}),\\&
\mu^{ab}_1=\lambda_1^{ab}+\delta^{ab}g_1({\bf x}),
\\&
\mu^{ab}_2=\tilde\lambda_2^a x^b+\tilde\lambda_2^b x^a-\delta^{ab}(2\lambda_3^c
x^c-g_2({\bf x})),
\\&
\mu^{ab}_3=(\varepsilon^{acd}\lambda_3^{cb}+ \varepsilon^{bcd}
\lambda_3^{ca})x^d+
\delta^{ab}g_3({\bf x}), \\&
\mu^{ab}_4=(x^a\varepsilon^{bcd}+x^b\varepsilon^{acd}) x^c\lambda^d_4+
\delta^{ab}g_4({\bf x}),
\\&\mu^{ab}_5=\delta^{ab}(r^2+g_5({\bf x}))+
k (x^ax^b-\delta^{ab}r^2),\\&
\mu^{ab}_6=\lambda_6^{ab}r^2-(x^a\lambda_6^{bc}+x^b\lambda_5^{ac})x^c-
\delta^{ab}(\tilde\lambda_6^{cd}x^cx^d-g_6({\bf x})),\end{split}\\\begin{split}
& \mu^{ab}_7=(x^a\lambda_7^b+x^b\lambda_7^a)r^2-4x^ax^b\lambda_7^c x^c+
\delta^{ab}(
\tilde \lambda_7^c x^cr^2+g_7({\bf x})),
\\&\mu^{ab}_8= 2(x^a\varepsilon^{bcd} +x^b\varepsilon^{acd})
\lambda_8^{dn}x^cx^n- (\varepsilon^{ack}\lambda_8^{bk}+
\varepsilon^{bck}\lambda_8^{ak})x^cr^2+
\delta^{ab}g_8({\bf x})\\&
\mu^{ab}_9=\lambda_9^{ab}r^4-2(x^a\lambda_9^{bc}+x^b\lambda_9^{ac})x^cr^2+
(4x^ax^b+\delta^{ab}r^2)\lambda_{9}^{cd}x^cx^d\\&+
\delta^{ab}(\tilde\lambda_{9}^{cd}x^cx^dr^2+
g_9({\bf x}))\end{split}\la{mmu2}
\end{gather}
where $r=\sqrt{x_1^2+x_2^2+x_3^2}$, $\lambda_n^{ab}=\lambda_n^{ba}, \tilde\lambda_n^{ab}=\tilde\lambda_n^{ba}$ and  $\lambda_n^a $  are
arbitrary parameters, and  $g_1,...,g_9$ are arbitrary functions of
$\bf x$.

Thus our classification problem is reduced to finding inequivalent solutions of equations
(\ref{mmmu1})
and (\ref{mmmu4}) where $\mu^{ab}$ are linear combinations of tensors (\ref{mmu1}), and generic
form of  functions $f$ and $V$ is specified
 in (\ref{f_V1})--(\ref{f_V3}).

The mentioned linear combinations are the fourth order polynomials in $x_a$  and include nine  arbitrary
functions and as many as  50 arbitrary parameters, and so in this stage the classification problems
looks huge indeed.
  Fortunately, for the systems whose inverse masses are specified in (\ref{IM1})-(\ref{IM3})  this
  problem can be reduced to the series of relatively
  simple subproblems corresponding to particular linear combinations of
  these tensors.

  \section{Scale invariant PDM systems}

   Let us start with the systems admitting three dimensional symmetry algebra isomorphic to  so(1,2).
   The corresponding Hamiltonians are specified by equations  (\ref{se}), (\ref{H}) and (\ref{f_V3}) while
   the related symmetries are given in (\ref{IM3}).
        The mentioned systems admit second order symmetry operators (\ref{Q}) provided equations
        (\ref{mmmu1})
and (\ref{mmmu4}) are satisfied. In particular these systems by definition should be invariant w.r.t.
the dilatation transformations whose generator $D$ is present in the list  (\ref{IM3}).

We will consider even a more generic problem. Namely, let us temporary forget about symmetries
generated by the shift generator $P_a$ and generator $K_a$ of the conformal transformations, and
solve the determining equations for the masses and potentials admitting only the dilatation symmetry.
Such problem has its own value and is an important subproblem of classification of PDM systems with
Lie symmetry groups  including  the dilatation as a subgroup.

In this case we have a bit more general forms of $f$ and $V$ than ones fixed in  (\ref{f_V3}), namely
\begin{gather}\la{f_V4}f= r^2F(\varphi,\theta),\quad V=V(\varphi,\theta)\end{gather}
where $F(.)$ and $V(.)$ are arbitrary functions, $\varphi$ and  $\theta$ are the Euler angles. After
finding all inequivalent symmetries for systems with the inverse masses and potentials specified in
(\ref{f_V4}) we will impose the additional conditions $\frac{\p f}{\p\theta}=0$ and    $\frac{\p
V}{\p\theta}=0$ and obtain the systems with $SO(1,2)$ symmetry. In addition, asking for the solutions
of the determining equations satisfying  $\frac{\p f}{\p \varphi}=0$ and    $\frac{\p V}{\p \varphi}=0 $
we come to the systems, admitting the two parametric Lie group including dilatations and rotations
around the third coordinate axis, etc.

  \subsection{Equivalence relations and reduction of the determining equations}

Changes of dependent and independent variables are called the equivalence tra nsformations provided
they keep the generic form of the differential equation (in our case of equation  (\ref{se})) up to the
changes of the explicit form of  arbitrary elements  (in our case functions $f$ and $V$). The set of the
equivalence transformations   includes equivalence groups extended by  some discrete elements.

In accordance with the results presented  in \cite{NZ}, the maximal continuous equivalence group of
equation  (\ref{se}) is C(3), i.e., the group of conformal transformations of the 3d Euclidean space. The
basis elements of the corresponding Lie algebra can be chosen  in the following form :
\begin{gather}\label{QQ}\begin{split}&
 P_{a}=p^{a}=-i\frac{\partial}{\partial x_{a}},\quad L_{a}=\varepsilon^{abc}x^bp^c, \\&
D=x_n p^n-\frac{3\ri}2,\quad K_{a}=r^2 p^a -2x^aD,\end{split}
\end{gather}
where $r^2=x_1^2+x_2^2+x_3^2$  and $p_a=-i\frac{\p}{\p x_a}.$ Operators $ P^{a},$ $ L^{a},$ $D$
and $ K^{a}$ generate shifts, rotations, dilatations and pure conformal transformations respectively.
The corresponding group transformations (whose explicit form can be found, e.g.,  in \cite{NZ}) keep
the generic form of equations  (\ref{se}), (\ref{H}) but can change the explicit form of  $f$ and $V$.

In addition to the invariance with respect to dilatation transformations the considered equations admit
the discrete inverse transformation:
 \begin{gather}\la{IT} x_a\to
\tilde x_a=\frac{x_a}{r^2},\quad \psi({\bf x})\to \tilde x^3\psi(\tilde{\bf x}), \ \tilde x=\sqrt{\tilde x_1^2+\tilde x_2^2+\tilde x_3^2}
\end{gather}
which acts on operators  (\ref{QQ}) in the following manner:
\begin{gather} P_a\to K_a,\quad K_a\to P_a,\quad L_a \to L_a,\quad D \to D\la{inv}\end{gather}

For the class of equations considered in the present section the equivalence group is reduced to the
direct product of the rotations group and dilatation transformations since $L_a$ and $D$ commute
with $D$ while the remaining operators  (\ref{QQ}) do not have this property.

In the following we will use the rotations and the inverse transformation  (\ref{IT}) for optimisation of
calculation.

Since the considered  systems by definition should be invariant w.r.t. the scaling transformations
(whose generator $D$ is present in the list  (\ref{IM3})), the related Killing tensors cannot include
linear combinations of all polynomials listed in  (\ref{mmu1}) but are reduced to homogeneous
polynomials.  In other words, the determining equations     (\ref{mmmu1})
and (\ref{mmmu4}) are reduced to the five  decoupled subsystems corresponding to the Killing  tensors
which are $n$-order homogeneous polynomials with $n=0, 1, 2, 3, 4$, and arbitrary functions
$g_1, g_2, ...,g_9$ should satisfy the following equations:
\begin{gather}\la{vp}x_ag({\bf x})_a=ng({\bf x}).\end{gather}
Moreover, since Hamiltonians
(\ref{H}) with arbitrary elements
(\ref{f_V3}) are invariant with respect to the inverse transformation (\ref{IT})
 we can restrict ourselves to the polynomials of order $n < 3$, since symmetries with $n$=3 and
 $n$=4 appears to be equivalent to ones with $n=1$ and $n=0$ correspondingly.

Thus it is sufficient to solve  determining equations (\ref{mmmu1})
and (\ref{mmmu4}) with the following versions of functions $\mu^{ab}$:
\begin{gather}\la{m00}\mu^{ab}=\tilde \mu^{ab}+\delta^{ab}g\end{gather}
where $g =g({\bf x})$ and
\begin{gather}\la{m0}\tilde\mu^{ab}=\lambda^{ab},\\\la{mm1}\begin{split}
\tilde\mu^{ab}=\lambda^a x^b+\lambda^b x^a-2\delta^{ab}\lambda^cx^c +\mu^ax^b+\mu^bx^a
+(\varepsilon^{acd}\lambda^{cb}+ \varepsilon^{bcd}
\lambda^{ca})x^d,\end{split}\\\la{mm2}\begin{split}&
\tilde \mu^{ab}=\kappa x^ax^b+(x^a\varepsilon^{bcd}+x^b\varepsilon^{acd})\lambda^dx^c+
\delta^{ab}\tilde\lambda^{cd}x^cx^d+
\lambda^{ab}r^2-(x^a\lambda^{bc}+x^b\lambda^{ac})x_c.\end{split}\end{gather}

Formula (\ref{mm1}) represents  tensor $\mu_2^{ab}$ from (\ref{mmu1}) with  slightly modernized
notations $\lambda_2^a=\lambda^a$ and $\tilde\lambda_2^a=\mu^a+\lambda^a$. We also omit the
sub indices for $\lambda^a$ and $\lambda^{ab}$.

 Let us note note  that the integrals of motion (\ref{Q}) corresponding to the Killing tensors (\ref{m00}) can be expressed via generators (\ref{QQ}):
\begin{gather}\la{ea1}Q=\lambda P_aP_b +\eta+P_agP_a,\\\la{ea2}Q=\lambda^a\{P_a,D\}+\lambda^{ab}\{P_a,L_b\} +\eta+P_agP_a\end{gather}
and
\begin{gather}\la{ea3}Q=\nu^{ab}(\{K_a,P_b\} +\{P_b,K_a\})+\tilde \lambda^{ab}Q^{ab},\end{gather}
where $\nu^{ab}=\lambda^{ab}+\varepsilon_{abc}\lambda_c,$ $\tilde \lambda^{ab}$ are arbitrary koefficients,  $Q^{ab}=P_cx_ax_bP_c$, and the symbols $\{.,.\}$ denote anticommutators.  Representations (\ref{ea1}),  (\ref{ea2}) and  (\ref{ea3}) correspond to tensors (\ref{m0}), (\ref{mm1})
and (\ref{mm2}) respectively. These representations are not unique since we cane indicate the following identities:
\begin{gather}\la{Id}\begin{split}&\{P_a,D\}+\varepsilon_{abc}\{P_b,L_c\}=2P_cx_aP_c, \\& \{L_a,L_b\}+\{P_a,K_b\}=2Q^{ab},\quad  a\neq b,\\&
\{P_1,K_1\}+\{ P_2,K_2\}+L_3^2=2Q^{33}.\end{split}\end{gather}

Notice also than any second order symmetry corresponding to $n=0$ and $n=1$ is accompanied by the
addition symmetry generated by the changes of variables  (\ref{IT}) and transformations (\ref{inv}).
\subsection{Evolution of the determining equations}
The first step of our analysis is to evolute determining equations (\ref{mmmu1}) for the inverse mass functions  and functions $g$ included to Killing tensors.

For the systems invariant w.r.t. the dilatation transformations function $f$ satisfies one more
condition
\begin{gather}\la{ff}x_af_a=2f\end{gather}
which is obviously correct in view of (\ref{f_V4}). However this condition  enables to reduce (\ref{mmmu4}) to the
following    {\it homogeneous} system of linear algebraic equations for derivatives   $f_a$:
\begin{gather}\la{NO}M^{ab}f_b=0\end{gather}
where
\begin{gather*}\begin{split}& M^{ab}=\tilde \mu^{ab}-g_ax_b,\\&  M^{ab}=\tilde \mu^{ab}-\lambda^ax^b-\mu^ax_b-g_ax_b\end{split}\end{gather*}
and
\begin{gather} \la{deq51}M^{ab}=\tilde \mu^{ab} -\lambda^{ac}x_cx_b-g_ax_b\end{gather}
for Killing vectors (\ref{m0}), (\ref{mm1}) and (\ref{mm2}) correspondingly.

Notice that  for the Killing vectors (\ref{m0}) and (\ref{mm1}) functions  $g({\bf x})$
can be expressed via $f$  by the following equations:

\begin{gather}\la{vp1}g({\bf x})=\frac1{2f}{-x_a M^{ab}f_b}\end{gather}
and
\begin{gather}\la{vp2}g({\bf x})=\frac1{f}{-x_a M^{ab}f_b}{f}\end{gather}
correspondingly, while for Killing vectors (\ref{mm2}) we have:
\begin{gather}\la{vp3}g({\bf x})=fG(g,\theta)\end{gather} where $ G(\varphi,\theta)$ is yet unknown function of Euler angles, satisfying the equation
\begin{gather}\la{vp4}G_\varphi=\frac1{f^2}(x_aM^{bc}f_c-x_bM^{ac}f_c)\end{gather}

Equations (\ref{vp1}) - (\ref{vp4}) are algebraic consequences of (\ref{vp}), (\ref{ff}) and (\ref{NO}). obtained by multiplication on $x_a$ and summing up with respect to the repeating index $a$.

Equation (\ref{NO}) admits nontrivial solution iff the determinant of the matrix whose entries are
$M^{ab}$
 is equal to zero. Thus we have to specify the admissible combinations of arbitrary constants nullifying
 this determinant and than find solutions of the corresponding equations (\ref{NO}) and
 (\ref{mmmu4}).
 For the Killing tensors presented in (\ref{m0})-(\ref{mm2}) the latter equation  is simplified to the
 following form
\begin{gather}\la{mmmu7}f\eta^a-\mu^{ab}V_b=0. \end{gather}

Thus our classification problem is reduced to solving the system of equations  (\ref{NO}) and (\ref{mmmu7})
for unknowns $f, g$ and $V$.

We will not present all the related routine  calculations whose details can be found in \cite{ANG}, but
restrict ourselves to one special case which is missing there.

 \subsection{Polynomial potentials}

 Consider  the most complicated case when the symmetry operator  is the reduced to the following
 bilinear form of generators of group C(3):
\begin{gather}\la{bc}Q=\mu\{K_1,P_1\}+\kappa\{K_2,P_2\}+2\eta\end{gather}
where $\mu$ and $\kappa$ are arbitrary coefficients, which, up to normalization, are supposed   to
satisfy the condition $\mu^2+\kappa^2=1$.

 The corresponding matrix $M$  is degenerated, and its nonzero  entries
take the following form:
\begin{gather*}M^{11}=\mu( x_3^2+x_2^2),\ M^{12}=-\kappa x_1x_2,\ M^{21}=-x_1x_2,\\
M^{22}=\kappa (x_1^2+ x_3^2), \ M^{31}=- \mu x_1x_3, \
M^{32}=-\kappa x_2x_3.\end{gather*}

The related equations (\ref{NO}) are solved by $f=x_3^2$, and the corresponding equations
(\ref{mmmu4}) take the following form:
\begin{gather}\la{new1}\begin{split}&\mu(x_2^2+x_3^2)V_1-\kappa x_1x_2V_2=x_3^2\eta_1,\\&
\kappa(x_1^2+x_3^2)V_2-\mu x_1x_2V_1=x_3^2\eta_2.\end{split}\end{gather}
 Notice that the third component of equations  (\ref{mmmu3}) in our case is a consequence of the
 system (\ref{new1}) since $\eta$ should satisfy the condition $x_a\eta_a=0$.

By definition potential $V$ should be scale invariant and so can be treated as a function of two scale
invariant variables
\begin{gather}\la{new27}y_1=\frac{x_1}{x_3},\ {\text{and}}\ y_2=\frac{x_2}{x_3}.\end{gather} The
system  (\ref{new1}) is compatible provided the following second order equation for $V$ is satisfied:
\begin{gather}\la{new2}(\mu y_2^2-\kappa y_1^2+\mu-\kappa)V_{ y_1y_2}+y_1y_2(\mu V_{y_1y_1}-\kappa V_{y_2
y_2})+3(\mu y_2V_{y_1}-\kappa y_1V_{y_2}) = 0.\end{gather}

The  system (\ref{new1}) can be easy solved for special combinations of parameters $a$ and $b$,
namely, $\mu =\kappa $ and $\mu =0$ (or $\kappa =0$ which is the same up to rotation).
However, to solve this system for $\mu\kappa(\mu-\kappa)\neq0$ a rather spectacular approach is requested.

In the latter case we can restrict ourselves to the parameters values satisfying the following
conditions:
 \begin{gather}\la{new20}\mu\kappa(\mu - \kappa)>0, \quad \mu^2+\kappa^2=1.\end{gather}
 It can be done without lost of generality up to normalization of the symmetry operator (\ref{bc}) and
 the equivalence transformations which are reduced to the rotations with respect to the third
 coordinate axis.

To simplify the related equation (\ref{new2}) it is convenient to use the following variables:
\begin{gather}\la{new22}x=\mu y_1^2-\kappa y_2^2+\mu\kappa(\mu-\kappa),\quad
y=\frac{\kappa y_1^2+\mu y_2^2+(\mu-\kappa)(\mu^2-\kappa^2)}{\sqrt{\mu\kappa(\mu- \kappa)}}\end{gather}
which reduce it to the following form:
\begin{gather}\la{new21}V_{yy}=yV_{xy}+xV_{xx}+2V_{x}=0.\end{gather}
By construction the latter equation has to have polynomial solutions which we find in the following
form:
\begin{gather}\la{new23}\begin{split}&V^{(s)}=y^s+(s-1)xy^{s-2}+
\frac{(s-2)(s-3)}{2}x^2y^{s-4}+...\\&
+\frac{(s-k)(s-k+1)(s-k+2)...(s-2k+1)}{k!}x^ky^{s-2k}+...\\&+\left(\delta\left(\frac{s-1}2\right)+1\right)x^{\frac{s-\delta}2}y^\delta\end{split}\end{gather}
where $\delta=1$ for even $s$ and $\delta=0$ for $s$ odd. In particular,
\begin{gather}\la{new24}\begin{split}&V^{(1)}=y,\\&V^{(2)}=y^2+x,\\&V^{(3)}=y^3+2xy,\\
&V^{(4)}=y^4+3xy^2+x^2,\\&V^{(5)}=y^5+4xy^3+3x^2y,\\&V^{(6)}=y^6+5y^4x+6y^2x^2+x^3
\end{split}\end{gather}where $x$ and $y$ have to be expressed via the initial variables $x_1, x_2$ and
$x_3$ by formulae (\ref{new22}) and (\ref{new27}).

Of course, a linear combination of generic polynomials (\ref{new23}) and their particular cases
presented in (\ref{new24}) also solves equation (\ref{new21}).
In addition, we can fix a multi parametric parametric solution which cannot be expressed via linear
combinations of polynomials (\ref{new23}):
\begin{gather}\la{new25}V=\frac{\alpha
x_3^2}{\kappa^2x_1^2+\mu^2x_2^2-
\kappa\mu x_3^2}.\end{gather}

Thus we find a countable set of  integrable PDM systems, admitting second order integrals of motion
The next step is to find the corresponding functions $\eta$. For any fixed potential $V$ enumerated in
(\ref{new24}) and (\ref{new25}) it can be easily done solving equations (\ref{new1}). In particular, for
potentials (\ref{new25}) we obtain
\begin{gather*}\eta=\frac{\kappa{x_1^2}+\mu{x_2^2}}
V.\end{gather*}

Notice that for some particular values of arbitrary parameters  the PDM systems with potentials
(\ref{new25}) have more extended symmetries which are indicated  in Items 3 and 4 of  Table 2.

In analogous way we can solve the  remaining inequivalent compatible systems (\ref{NO}) and (\ref{mmmu4}) which in fact are more easy to handle. To save a room we will not present the calculation details here since they can be found in \cite{ANG}.

Thus we have classified the PDM systems which are scale invariant and admit second order integrals
of motion. These integrals belong to one out of two subclasses. The first of them includes integrals which belong to the enveloping algebra of the conformal algebra c(3) up to constant  terms including functions of $\bf x$. The other subclass includes integrals of motions which do not belong to this enveloping algebra.
We have found all of them, but in the following Tables 1 and 2 just the systems  belonging  to the first subclass are presented. In contrary, Table 3 collects the systems which belong to the second class and are invariant with respect to the algebra so(1,2).   The presented list of
PDM systems admitting second order integrals of motion is complete up to rotation transformations.

In the tables  $F(.), G(.)$ and  $R(.)$ are arbitrary functions of the arguments specified in brackets,
$V^{(s)}$ are polynomials (\ref{new23}), $c, c_1, c_2, \mu$ and $\nu$ are arbitrary real parameters
$\varphi$ and $\theta$ are Euler angles, $ r^2=x_1^2+x_2^2+x_3^2,\ \tilde r^2=x_1^2+x_2^2, $ $
P_a, K_a, D$ and $L_3$ are operators defined in (\ref{QQ}), and the summation is imposed over the
repeating indices  $a$ by values 1, 2 and 3. The symbol $\{A,B\}$ denotes the anticommutator of
operators  $A$ and $B$, i.e.,
$\{A,B\}=AB+BA.$

 \begin{center}Table 1. Inverse masses, potentials  and the related integrals of motion defined up to
 arbitrary functions.\end{center}
\begin{tabular}{c c c c}

\hline

\vspace{1.5mm}

 No&$f$&$V$&\text{Integrals of motion}\\
\hline\\
1\vspace{1.5mm}&$\tilde r^2F(\theta)$&$G(\varphi)F(\theta)+R(\theta)$&$\begin{array}{c}
L_3^2+G({\varphi})\end{array}$\\

2\vspace{1.5mm}&$ r^2F(\theta)$&$cF(\theta)\varphi+G(\theta)$&$\begin{array}{c}
\{L_3,D\}+2c\ln(r)\end{array}$\\

3\vspace{1.5mm}&$\tilde
r^2F(\varphi)$&$F(\varphi)G(\theta)+R(\varphi)$&$\{P_3,K_3\}+2G(\theta)$\\

4\vspace{1.5mm}&$\tilde r^2F(\varphi)$& $c
F(\varphi)\frac{x_3}r+G(\varphi)$&$\begin{array}{c}\{P_3,D\}-\frac{2c}{r}, \ 
\{P_3,K_3\}+\frac{2cx_3}{r},\\ 
\{K_3,D\}-{2c}{ r}-12x_3 \end{array}$\\

5\vspace{1.5mm}&$\tilde r^2F(\varphi)$&$c\frac{\tilde
r^2}{x_3^2}F(\varphi)+R(\varphi)$&$\begin{array}{c}P_3^2+\frac{c}{x_3^2},\ K_3^2+3(4x_3^2-\tilde r^2)+\frac{cr^4}{x_3^2},\\
\{P_3,K_3\}+\frac{2c\tilde r^2}{x_3^2}\end{array}$\\
6\vspace{1.5mm}&$\tilde r^2$&$G(\theta)$&$\begin{array}{c} \{P_3,K_3\}+2G(\theta), \
L_3\end{array}$\\

7\vspace{1.5mm}&$x_3^2$& $\frac{x_3^2}{\tilde r^2}F(\varphi)$&$\begin{array}{c}
P_1^2+P_2^2+\frac{F(\varphi)}{\tilde r^2} , \  K_1^2+K_2^2+3(3\tilde r^2-2x_3^2)+\frac{F(\varphi)r^4}{\tilde r^2},\\
L_3^2+F(\varphi) \end{array}$\\

8\vspace{1.5mm}&$x_3^2$&$
V^{(s)}(\varphi,\theta)$&$\begin{array}{c}b\{P_1,K_1\}+a\{P_2,K_2\}+2\tilde \eta
\end{array}$\\

\hline \hline
\end{tabular}

\subsection{Algebraic structure of integrals of motion}
It is an element of common knowledge that the commutator of integrals of motion is the integral of
motion too. In other words integrals of motion form a Lie algebra which, however, can be infinite
dimensional. Indeed, a commutator of $n$ order (in our case second order) differential operators is the
operator whose order is generally speaking $2n-1$. The next commutator will have the order $3n-2$,
etc., and   the discussed algebra can include infinite number of integrals of motion of arbitrary order.

However, for some special  symmetries the algebra of integrals of motion appears to be finite
dimensional. First, these integrals can simple commute. Secondly, the well known example is  the
Laplace-Runge-Lenz vector which form the algebra so(4) provided the representation space of this
algebra is the set of solutions of Schr\"odinger equation for the Hydrogen atom and some more general
quantum mechanical systems \cite{N2}.

For the systems considered in the above the algebras of integrals of motion are infinite dimensional,
but their structure is rather transparent. Namely, let $Q_1, Q_2, ..., Q_n$ are second order integrals
of motion for one of the system. Then they satisfy the  following generic commutation relations:
\begin{gather}\la{koko} [Q_a, Q_b]=c^{ab}_kQ_kQ^{(1)}\end{gather}
where $C^{ab}_c$ are structure constants and $Q^{(1)}$ is the dilatation operator specified in
(\ref{IM3}) or some other first order integral of motion. However, this rule has one exception.

By definition  integrals of motion  commute with the Hamiltonian $H$. Thus $H$ and $Q^{(1)}$ have
the same set of eigenfunctions, and relations (\ref{koko}) specify the Lie algebra whose representation
space is an eigenvector of $Q^{(1)}$.

The structure constants $c^{ab}_k$ can be easily found by the direct calculation. We will not do this
routine job it for all sets of integrals of motion presented in the tables, but restrict ourselves to the
systems presented in Table 1.

Let us denote the integrals of motion presented in Table 1 as
\begin{gather}\la{ss}Q_1=\{P_3,K_3\}-2D^2+\frac{2cx_3}{r}, \ Q_2=\{P_3,D\}-\frac{2c}{r}, \
Q_3=\{K_3,D\}-{2c}{ r},\,\\Q_4=P_3^2+\frac{c}{x_3^2},\ Q_5=K_3^2+\frac{cr^4}{x_3^2},\
Q_6=\{P_3,K_3\}+\frac{2c\tilde r^2}{x_3^2},\\Q_7=P_1^2+P_2^2+\frac{F(\varphi)}{\tilde r^2} , \
Q_{8}=K_1^2+K_2^2+\frac{F(\varphi)r^4}{\tilde r^2},\\ Q_{9}=L_3^2+H+F(\varphi)\end{gather}
where $H$ is Hamiltonian and $D^2$ is the squared dilatation. We add the latter obvious symmetries into (\ref{ss}) to simplify the coupling constants. Calculating their commutators we specify the following algebraic
structures:
\begin{gather}\la{ko1}\begin{split}&[Q_7,Q_{8}]=2\ri Q_{9}D, \
[Q_7,Q_{9}]=[Q_{8},Q_{9}]=0,\end{split}\\\la{ko2}\begin{split}&[Q_1,Q_2]=-2\ri Q_2D, \ [Q_1,Q_3]=2\ri
Q_3D,\ [Q_3,Q_2]=2\ri Q_1D,\\&[Q_4,Q_5]=4\ri Q_6D, \ [Q_4,Q_6]=2\ri Q_4D, \ [Q_5,Q_6]=-2\ri
Q_5D.\end{split}\end{gather}

The presented commutators are proportional to the dilatation generator. Acting by the operators in the
l.h.s and r.h.s. on the eigenvectors of this generator we recognize that relations (\ref{ko1}) specify the
Heisenberg  algebra  while relations (\ref{ko2}) specify the Lie algebras isomorphic to so(1,2).

An example of the situation when the commutator of the second order integrals of motion is not
proportional to $D$ but to another first order symmetry (namely, $L_3$) can be found in Item 1 of
Table 2. The only case when such commutators are not proportional  to the first order symmetry is
present in Item 8 of the same table.
\vspace{1mm}

 \begin{center}Table 2.  Inverse   masses, potentials  and  integrals of motion defined up to arbitrary
 coefficients.\end{center}

 \vspace{2mm}

\begin{tabular}{c c c c}
\hline
\vspace{0mm}No&$f$&$V$&\text{Integrals of motion}\\
\hline
1\vspace{0.5mm}&$r^2$&$c\frac{r^2}{x_3^2}$&$\begin{array}{c}
L_2^2-L_1^2+c\frac{x_2^2-x_1^2}{x_3^2},\\ \{L_1,L_2\}+2c\frac{x_1x_2}{x_3^2}, \ L_3\end{array}$\\

2\vspace{1.5mm}&$x_3^2$&$\begin{array}{c}\frac{\alpha
x_3^2}{\kappa^2x_1^2+\mu^2x_2^2-
\kappa\mu x_3^2}\end{array}$&$\begin{array}{c}\mu\{P_1,K_1\}+\kappa
\{P_2,K_2\}
+\frac{\mu^2 x_1^2+\kappa{x_2^2}}{x_3^2}
V\end{array}$\\

3\vspace{1.5mm}&$x_2^2$&$c_1\frac{x_2^2}{x_3^2}+c_2\frac{x_1}{\tilde
r}$&$\begin{array}{c}\{P_1,D\}-\{P_3,L_2\}+c_2\frac{1}{\tilde r}-2c_1\frac{x_1}{x_3^2},\\
\{K_1,D\}-\{K_3,L_2\}-15x_1+c_2\frac{r^2}{\tilde r}-2c_1\frac{r^2x_1}{x_2^2},\end{array}$\\

4\vspace{1.5mm}&$x_3^2$&$c_1\frac{x_3^2x_2}{\tilde r
x_1^2}+c_2\frac{x_3^2}{x_1^2}$&$\begin{array}{c}\{L_3,P_1\}
+2c_2\frac{x_2}{x_1^2}+c_1\frac{2x_2^2+x_1^2}{\tilde r x_1^2},\\\{L_3,K_1\}+3x_2
+2c_2\frac{x_2r^2}{x_1^2}+c_1\frac{(2x_2^2+x_1^2)r^2}{\tilde r x_1^2},\\ K_1^2+K_2^2+\frac{r^4}{ x_3^2}V+3(3\tilde r^2-2x_3^2), \\ L_3^2+c_1\frac{\tilde r x_2}{ x_1^2}+c_2\frac{\tilde
r^2}{x_1^2}, \ P_1^2+P_2^2+\frac{1}{
x_3^2}V \end{array}$\\

5\vspace{1.5mm}&$x_3^2$&$c\frac{x_3^2}{x_1^2}$&$
\begin{array}{c}L_3^2+c\frac{\tilde r^2}{x_1^2}, \ P_2, \ K_2,\\\{L_3,P_1\} +2c\frac{x_2}{x_1^2}, \
\{L_3,K_1\} -3x_2+2c\frac{x_2r^2}{x_1^2}, \ \\ P_1^2+\frac{c}{x_1^2},\  K_1^2+3(5x_1^2-r^2)+\frac{cr^4}{x_1^2},  \{P_1,K_1\}+\frac{2cr^2}{x_1^2}, \
\end{array}$\\

6\vspace{1.5mm}&$x_3^2$&$c_1\frac{x_3^2}{x_1^2}+c_2\frac{x_3^2}{x_2^2}$&$
\begin{array}{l}\ P_1^2+\frac{c_1}{x_1^2},\  P_2^2+\frac{c_2}{x_2^2}, \   K_1^2+3(5x_1^2-r^2)+\frac{c_1r^4}{x_1^2},\\
K_2^2+3(5x_2^2-r^2)+\frac{c_2r^4}{x_2^2},\\  \{P_1,K_1\}+\frac{2c_1r^2}{x_1^2}, \ \{P_2,K_2\}+\frac{2c_2r^2}{x_2^2}
\end{array}$\\

7\vspace{1.5mm}&$x_3^2$&$c\frac{\tilde r^2}{r^2}$&$\begin{array}{c}
\{K_1,P_1\}-\{K_2,P_2\}+2c\frac{x_1^2-x_2^2}{r^2},\\\{K_1,P_2\}+\{K_2,P_1\}+4c\frac{x_1x_2}{r^2}, \
L_3\end{array}$\\

8\vspace{1.5mm}&$x_3^2$& $c\frac{x_3^2}{\tilde r^2}$&$\begin{array}{c} P_1^2+P_2^2+\frac{c}{\tilde
r^2},\ L_3 , \\ K_1^2+K_2^2+3(3\tilde r^2-2x_3^2)+\frac{cr^4}{\tilde r^2} \end{array}$\\

9\vspace{1.5mm}&$\tilde r^2$&$c_1e^{-2\varphi}\frac{r^2+x_3^2}{\tilde
r^2}+c_2e^{-\varphi}\frac{x_3}{\tilde r}$&$\begin{array}{l}\{P_3,(L_3+
D)\}+c_1e^{-2\varphi}\frac{x_3}{\tilde r^2}+c_2e^{-\varphi}\frac1{\tilde r},\\\{K_3,(L_3+
D)\}-12x_3+c_1e^{-2\varphi}\frac{r^2 x_3}{\tilde r^2}+c_2e^{-\varphi}\frac{r^2}{\tilde r}\end{array}$\\

10\vspace{1.5mm}&$\tilde r^2$&$c_1e^{2\varphi}\frac{r^2+x_3^2}{\tilde
r^2}+c_2e^\varphi\frac{x_3}{\tilde r}$&$\begin{array}{l}\{P_3,(L_3- D)\}-c_1e^{2\varphi}\frac{x_3}{\tilde
r^2}-c_2e^{\varphi}\frac1{\tilde r},\\\{K_3,(L_3- D)\}+12x_3-c_1e^{2\varphi}\frac{r^2x_3}{\tilde
r^2}-c_2e^{\varphi}\frac{r^2}{\tilde r}\end{array}$\\

11\vspace{1.5mm}&$\tilde r^2$& $c\frac{x_3}r$&$\begin{array}{c} \{K_3,D\}-12x_3-{c}{ r},\
L_3\\\{P_3,K_3\}+2c\frac{x_3}{ r},\ \{P_3,D\}-\frac{c}{r}\\
\end{array}$\\

12\vspace{1.5mm}&$\tilde r^2$&$c\frac{\tilde
r^2}{x_3^2}$&$\begin{array}{c}\{P_3,K_3\}+2c\frac{\tilde r^2}{x_3^2},  \ L_3, \ P_3^2+\frac c{x_3^2},\\
K_3^2+3(5x_3^2-r^2)+c \frac{r^4}{x_3^2} \end{array}$\\
\hline\hline
\end{tabular}

\vspace{3mm}

\section{PDM systems invariant with respect to algebra $so(1,2)$}

The results obtained in Section 4 can be generalized  to the cases  of PDM systems invariant w.r.t.
multi parametric  symmetry groups including the subgroup of dilatations. In this section we specify
the systems which are invariant with respect to the Lie group isomorphic to SO(1,2) and admit second
order integrals of motion.
The generic form of the corresponding Hamiltonians and the related first order symmetries are fixed in
(\ref{f_V3}) and (\ref{IM3}).

The considered systems are invariant with respect to dilatations, and so the related inverse masses
and potentials should be present in Tables 1 and 2. Thus our task is to select such systems   specified in
these tables with at least for some particular arbitrary functions or parameters are invariant with
respect to shifts and dilatations generated by operators $P_3$ and $K_3$ presented in  (\ref{IM3}). In
fact it is sufficient to ask for the symmetries with respect to the shifts since symmetry with respect to
the conformal transformations is a consequence of the symmetry with respect to shifts and dilatations,
see equations (\ref{IT}) and (\ref{inv}).

In other words everything we need is to to select such potentials and inverse masses presented in the
Tables 1 and 2 which satisfy  the additional conditions
\begin{gather}\la{an}\frac{\p f}{\p x_a}=0, \ \  \frac{\p V}{\p x_a}=0\end{gather}
for some fixed $a$ not necessary equal to 3. Whenever conditions (\ref{an}) are satisfied, we can
transform it to the case $a=3$ by a rotation transformation.

In addition, it is necessary to consider second order symmetries which do not belong to the enveloping algebra of c(3). The related Killing tensors initially include arbitrary functions $ g({\bf x})$ which are connected with the inverse mass functions $f$ via relations (\ref{vp1})-(\ref{vp4}).

Let us consider the most important symmetry which is possessed by all analysed systems. It has the following generic form:
\begin{gather}\la{L3}Q=L_3^2 +P_a\phi P_a+\eta\end{gather}
where $L_3=-\ri(x_1\frac{\p}{\p x_2}-x_2\frac{\p}{\p x_1}),\ P_a=-\ri \frac{\p}{\p x_a}$, $\phi $ and $\eta$  are unknown functions of $\bf x$.

 The corresponding matrix $\tilde \mu^{ab}$  is degenerated, and its nonzero  entries
are:
\begin{gather}\la{mumu}\tilde \mu^{11}=x_2^2, \quad \mu^{22}=x_1^2,\quad \tilde \mu^{12}=\tilde \mu^{21}=-x_1x_2.\end{gather}
To obtain these entries from formula (\ref{mm2}) we have to set $\lambda^{11}=\tilde\lambda^{11}=\tilde\lambda^{22}= \tilde\lambda^{33}=1$ while the remaining entries $\lambda^{ab}$ and $\tilde\lambda^{ab}$ are equal to zero.

In accordance with (\ref{mumu}) the related equations (\ref{NO}) are reduced to the following system:
\begin{gather}\la{s1}\begin{split}&x_2f_\varphi-gf_1+fg_1=0,\\&x_1f_\varphi+gf_2-fg_2=0,\\&gf_3-fg_3=0
\end{split}\end{gather}
where $f_\varphi=\frac{\p f}{\p \varphi}$ and $f_a=\frac{\p f}{\p x_a}$.

Since we are dealing with the Killing tensor quadratic in $x_a$ functions $f$ and $\varphi$ are connected by relation (\ref{vp4}) which reduce equations (\ref{s1}) to the following relations: \begin{gather}\la{s2}f_\varphi=f^2g_\varphi, \quad g_\theta=0\end{gather}
whose generic (up to constant multiplier) solution is
\begin{gather}\la{s3}f=\frac{\tilde r^2}{F(\varphi)+G(\theta)}.\end{gather}

Substituting (\ref{vp4}) and (\ref{s3}) into equation (\ref{mmmu7}) and integrating it we obtain
\begin{gather}\la{s4}V=\frac{M(\theta)+
N(\varphi)}{F(\varphi)+G(\theta)}, \eta= -F(\varphi)V+G(\theta).\end{gather}
The corresponding Hamiltonian commutes with $P_3$ provided functions $G(\theta)$ and $M(\theta)$ are constants, and we come to the results connected in Item 1 of Table 3, where we use the notations
\begin{gather}\la{NN}({F({\bf x})}\cdot H)=p_aF({\bf x})fp_a+ F({\bf x})V.\end{gather}

For some particular functions $F(\varphi)$ and $N(\varphi)$ the related Hamiltonians admit additional symmetries presented in Items 2-7 of the mentioned table.They can be calculated in analogy with the above.

%\vspace{2mm}
%\newpage

Table 3. Inverse masses and potentials for  equations admitting symmetry algebra $so(1,2)$

\vspace{3mm}

\begin{tabular}{c c c c}

\hline

\vspace{2mm}

 No&$f$&$V$&\text{Integrals of motion}\\
\hline\\

1\vspace{1.5mm}&$ \tilde r^2F(\varphi)$&$ F(\varphi)G(\varphi)$&$\begin{array}{c}
L_3^2- (\frac1{F(\varphi)}\cdot H)+G(\varphi)\end{array}$\\

2\vspace{1.5mm}&$x_2^2$&$F(\varphi)$&$\begin{array}{c} \{K_1,P_1\}-2L_2^2-2F(\varphi),\\
L_3^2- (\frac{\tilde r^2}{x_2^2}\cdot H)+\frac{x_2^2}{\tilde r^2}F(\varphi)\end{array} $\\

3\vspace{1.5mm}&$\frac{x_1^2\tilde r}{c_1 \tilde r +c_2 x_2}$&$\frac{c_3 \tilde r+c_4 x_2}{c_1 \tilde r +c_2 x_2}$&$\begin{array}{c} \{P_2,D\}+\{P_3,L_1\}+2\left(\frac{c_2}{\tilde r}\cdot H\right)-2\frac{c_4}{\tilde r} \\
\{K_2,D\}+\{K_3,L_1\}+2\left(\frac{c_2r^2}{\tilde r}\cdot H\right)-2\frac{c_4r^2}{\tilde r}-15x_2,\\L_3^2-\left(\frac{\tilde r(c_1 \tilde r +c_2 x_2)}{x_1^2 }\cdot H\right)+\frac{\tilde r(c_3 \tilde r+c_4 x_2)}{x_1^2}\end{array}$\\

4\vspace{1.5mm}&$\tilde r^2+\varepsilon x_1 \tilde r$&$\frac {c_1x_1+c_2 \tilde r}{\tilde r-\varepsilon x_1}$&$
\begin{array}{c}  \{P_2,D\}+\{P_3,L_1\}-2\left(\frac1{\tilde r}\cdot H\right)-2\frac{c_1}{\tilde r} ,\\

\{K_2,D\}+\{K_3,L_1\}-2\left(\frac{r^2}{\tilde r}\cdot H\right)-2\frac{c_1r^2}{\tilde r}-15x_2,

\\L_3^2- (\frac{\tilde r}{\tilde r+\varepsilon x_1}\cdot H)+\frac{(c_1x_1+c_2 \tilde r)\tilde r}{x_2^2}
\end{array}$\\

5\vspace{1.5mm}&$\frac{x_1^2 x_2^2 }{c_1 x_2^2+c_2x_1^2}$&$\frac{c_4 x_2^2+c_3x_1^2}{c_1 x_2^2+c_2x_1^2}$&$\begin{array}{c}P_1^2-\left(\frac{c_1}{x_1^2}\cdot H\right)+\frac{c_4}{  x_1^2}, P_2^2-\left(\frac{c_2}{x_2^2}\cdot H\right)+\frac{c_3}{ x_2^2},\\

K_1^2-\left(\frac{c_1r^4}{x_1^2}\cdot H\right)+\frac{c_4 r^4}{  x_1^2}+3(5x_1^2-r^2), \\ K_2^2-\left(\frac{c_2r^4}{x_2^2}\cdot H\right)+\frac{c_4r^4}{ x_2^2}+3(5x_2^2-r^2),\\

L_3^2- (\frac{\tilde r^2(c_1 x_2^2+c_2x_1^2)}{x_1^2x_2^2}\cdot H)+\frac{\tilde r^2(c_4 x_2^2+c_3x_1^2)}{x_1^2x_2^2}\end{array}$\\

6\vspace{1.5mm}&$x_2^2$&$c\frac{x_2^2}{x_1^2}$&$
\begin{array}{c} L_3^2+c\frac{\tilde r^2}{x_1^2}, \\\{L_2,P_1\}+2c\frac{x_3}{x_1^2}, \ \{L_2,K_1\} +2c\frac{x_3r^2}{x_1^2}-3x_3, \ \\ \   \{P_1,K_1\}+\frac{2cr^2}{x_1^2}, \ P_1^2+\frac{c}{x_1^2},\  K_1^2+\frac{cr^4}{x_1^2}+3(5x_1^2-r^2)
\end{array}$\\

7\vspace{1.5mm}&$\tilde r^2$&$c$&$P_3, \ L_3,\ D, \ K_3$\\

8\vspace{1.5mm}&$x_3^2$&$c$&$P_1,\  P_2, \ K_1,  \ K_2, D, L_3$\\

\hline \hline
\end{tabular}

\vspace{3mm}

The last two items of Table 3 include systems which admit only the first order integrals of motion
and their bilinear combinations.

\section{PDM systems invariant w.r.t. the rotation group}

 Let us discuss the rotationally invariant systems which are specified in equations (\ref{se}), (\ref{H})
 and (\ref{f_V1}), and present such of them which admit second order integrals of motion.

 In contrast with the scale invariant systems classified in Section 4 in this case we cannot decouple
 the determining equations with respect to the order in $\bf x$ of the related Killing tensors. However,
 these equation are reduced to systems of {\it ordinary} differential equations with well  defined
 tensorial properties. And it is possible (and necessary) to make the another type of decoupling
 corresponding to  the scalar, vector,  and  tensor integrals of motion. The related Killing tensors looks as follows:
 \begin{gather}\la{m1}\mu^{ab}=\mu^{ab}_5,\\
  \la{m2}\mu^{ab}=\nu \mu^{ab}_3+ \lambda\mu^{ab}_4,
  \\\la{m3}\mu^{ab}=\mu^{ab}_8,\\
\la{m4}\begin{split}&
\mu^{ab}=\nu\mu^{ab}_2+\mu\mu^{ab}_4=\nu(\lambda^a x^b+\lambda^b x^a-2\delta^{ab}\lambda^c x^c)\\&+\mu((x^a\lambda^b+x^b\lambda^a)x^2-4x^ax^b\lambda^c x^c+2\delta^{ab}
x^2\lambda^c x^c)+\delta^{ab}
\lambda^c x^c g(r),\end{split}\\\la{m5}
\begin{split}&\mu^{ab}=\nu\lambda^{ab}+
 \omega\left(\lambda^{ab}x^2-(x^2\lambda^{bc}+x^b\lambda^{ac})x^c
 -2\delta^{ab}\lambda^{cd}x^cx^d\right)\\& +
 \mu\left(\lambda^{ab}x^4-2(x^a\lambda^{bc}+x^b\lambda^{ac})x^cx^2+
 4x^ax^b\lambda^{cd}x^cx^d\right)+\delta^{ab}\lambda^{cd}x^cx^d g(r).
 \end{split}
\end{gather}
where $\mu^{ab}_n$  are tensors represented in (\ref{mmu2}).

Tensors (\ref{m1}), (\ref{m2}) and (\ref{m3}) correspond to the scalar, pseudovector and pseudotensor integrals of motion respectively. Notice that to keep  the correct transformation properties of the integrals of motion with respect to rotations arbitrary functions $g_n$ present in (\ref{mmu2}) should be reduced to zero for the case of pseudo vectors (tensors) and to functions of $r$ for the true ones.

The scalar integral is the squared  orbital momentum and is admitted by any of the considered systems. The pseudovector and pseudotensor integrals are not admitted by any system. The technical reason of the latter situation is just the  triviality of the arbitrary functions $g_n$.

For rotationally invariant inverse masses and potentials (\ref{f_V1}) the determining equations (\ref{mmmu1}) and (\ref{mmmu4})   are reduced to the following forms:
 \begin{gather}\la{m6}(\mu^{nn}_a+2\mu^{na}_n)f=
10\mu^{an}x_nf'\end{gather}
and
\begin{gather}\la{m7}f\eta^a-2\mu^{ab}x_bV'=0. \end{gather}
where $f'=\frac{\p f}{\p r^2}$ and  $V'=\frac{\p V}{\p r^2}.$

Let us consider the tensor integrals of motion generated by the Killing tensors (\ref{m5}). Substituting  (\ref{m5}) into (\ref{m6}) we come to the following system:
\begin{gather}\la{m8}\begin{split}& (g+\mu r^2)f'-(2\mu  +g)f=0,\\&(\mu r^4-\nu)g'+2(g-\mu r^2)f=0\end{split}\end{gather}
whose generic solutions are
\begin{gather}\la{m9}f=c_1 r^4, \quad g=0\end{gather}
and
\begin{gather}\la{m8}f =\frac{( r^4-\nu)^2}{c_1(\mu r^4+\nu) +c_2 r^2}, \quad g=-\frac{c_2(\mu r^4+\nu)+4\mu\nu c_1 r^2 }{c_1(\mu r^4+\nu) +c_2 r^2} \end{gather}
 where $c_1$ and $c_2$ are integration constants. Substituting (\ref{m9}) and (\ref{m8}) into (\ref{m7}) we easily find the corresponding potentials $V$ and functions $\eta$ which are presented in Items 4-6 of Table 4. Notice that the zero value of arbitrary constant $c_1$ correspond to the  special solution  for $V.$

 In analogous way we find the solutions of the determining equations (\ref{m6}) and (\ref{m7}) for the vector integrals of motion.

 \vspace{2mm}

 Table 4. Inverse masses and potentials for  equations admitting symmetry algebra $so(3)$

\vspace{4mm}

\begin{tabular}{c c c c}
\hline
No&$f$&$V$&\text{Integrals of motion}\\
\hline\\

1.&\vspace{2mm}$(\mu r^2-\nu)^2$&$\frac{\alpha (\mu r^2+\nu)}r$&$Q_a=\varepsilon_{abc}\{(\nu P_b+ \mu K_b) ,L_c\}+\frac{ x_a(\alpha-6r)}r$\\

2.&\vspace{2mm}$\frac{r(\mu r^2-\nu)^2}{\kappa r-2 (\mu r^2+\nu)}$&$\frac{\alpha r}{\kappa  r-2 (\mu r^2+\nu)}$&$\begin{array}{c}Q_a=\varepsilon_{abc}\{(\nu P_b+ \mu K_b) ,L_c\}\\-(\frac{ x_a}{r}\cdot H)-\frac{\alpha x_a}{ 2 (\mu r^2+\nu)-\kappa  r}-6x_a\end{array}$\\

3.&$(\mu\tilde \mu x^4-\nu\tilde\nu)^2$&$\frac{\alpha r^2}{(\mu\tilde \mu r^2+\nu\tilde\nu)^2}$&$\begin{array}{c}Q_{ab}=\{\mu K_a+\nu P_a,\tilde\mu K_b+\tilde\nu P_b\}\\+6\mu\tilde \mu (5x_ax_b-\delta^{ab}r^2)+\frac{2\alpha x^ax^b}{(\mu\tilde \mu x^2+\nu\tilde\nu)}\end{array}$\\

4.&\vspace{2mm}$\frac{(\mu r^4-\nu)^2}{r^2}$&$\frac{\alpha (\mu r^4+\nu)}{r^2}$&$\begin{array}{c}Q_{ab}=\frac\mu2\{K_a,K_b\} +\nu P_aP_b \\-\left(\frac{x_ax_b\mu(r^4+\nu)}{( \mu r^4-\nu)^2}\cdot H\right)\\+3\mu(5 x_ax_b -\delta^{ab}r^2)-\frac{4\alpha r^2x_ax_b}{(\mu r^4-\nu)^2}\end{array}$\\

5.&\vspace{2mm}$\frac{(\mu r^4-\nu)^2}{\mu r^4+\kappa r^2+\nu}$&$\frac{\alpha r^2}{\mu r^4+\kappa r^2+\nu}$&$\begin{array}{c}
Q_{ab}=\frac\mu2\{K_a,K_b\} +\nu P_aP_b+\\-\left(\frac{x_ax_br^2\mu(\kappa r^2+4\nu)}{(\mu r^4-\nu)^2}\cdot H\right)\\+3\mu(5 x_ax_b -\delta^{ab}r^2)-\frac{\alpha  x_ax_b(\mu r^4+\nu)}{(\mu r^4-\nu)^2}
\end{array}
$\\
6.&$(\mu x^4-\nu)^2$&$\alpha$&$\begin{array}{c}\mu K_a+\nu P_a\end{array}$\\

\vspace{2mm}
\\
\hline\hline \end{tabular}

\vspace{3mm}

 In Table 4 the Greek letters  denote arbitrary real
 coefficients. Any of them (except $\alpha$)   either can be  normalized to $\pm 1$ or be equal to zero. The last  of Table 4 includes systems which admit only the first order integrals of motion
and their bilinear combinations.

Notice that the systems presented in Items 1 and 2 of Table 4 are St\"ackel equivalent between themselves. The same is true for the systems fixed in Items 3, 4 and 5.
 We remind that the St\"ackel transform consists in the
multiplication of the Hamiltonian by the inverse potential (i.e., $H-E\to (V\cdot H)-\frac{E}V)$ where the operation  $(V\cdot H)$ should be treated in the sense defined by equation (\ref{NN}))  combined with the conformal
transformations and changing the roles played by coupling constants and Hamiltonian eigenvalues.
Moreover, at any step the potentials  can be added by a constant.

\section{PDM systems invariant w.r.t. the 2d Euclid  group E(2)}

Thus we have specified inequivalent superintegrable systems invariant with respect groups $SO(1,2)$
and $SO(3)$. The  last task is to describe the systems invariant with respect to the Euclid group
$E(2)$, whose inverse masses and potentials have the  form (\ref{f_V2}).

Since the arbitrary elements do not depend on $x_1$ and $x_2$ we can make rather restrictive a priori predictions about the possible second order integrals of motion.

Indeed, let $Q$ be an integral of motion.  By definition $P_a$
with $a=1, 2$  are integrals of motion too, the same is true for the commutators $[P_a,Q]$,
$[P_a,[P_b,Q]$ and $[P_a,[P_b,[P_c,Q]]$.  Since the Killing tensors are fourth order
polynomials in $\bf x$, and the derivation of the Killing tensor w.r.t. $x_a$ is again the Killing tensor,
we conclude that any second order symmetry induces the symmetry  generated by $\mu_0^{ab}$ (refer
to (\ref{mmu1}), i.e.,
\begin{gather}\la{a0}Q=P_a(\lambda_{ab}+\delta_{ab}G({\bf x}))P_b+\eta.\end{gather}

We can specify the following qualitatively different versions of coefficients $\lambda^{a b}$ in formula (\ref{a0}):
\begin{gather}\la{a1}\lambda^{3\mu}\neq0, \ \lambda^{\mu\nu}=\lambda^{33}=0, \  \mu, \nu=1,2,\\ \la{a2}
 \lambda^{3\mu}=0,  \text{\  some of coefficients \ } \lambda^{\mu\nu} \text{\ or \ } \lambda^{33} \text{\ are nontrivial} \end{gather}

 In the case (\ref{a1})  the systems admitting second order constants of motion necessary admit the following constants
 \begin{gather}\la{a3} Q^{31}=P_3P_1+\eta^{31}, \ \ Q^{32}=P_3P_2+\eta^{32},\end{gather}
moreover, both of them, in view of the symmetry with respect to rotations around the third coordinate axis.

In the case (\ref{a2}) the related integrals of motion  are trivial since the  Hamiltonians considered in this section commute with them by definition. It is evident for $Q^{\mu\nu}=P_{\mu}P_{\nu}+\eta^{\mu\nu}$ with $\mu, \nu<3$. On the other hand, $Q ^{33}=P_3^2+\eta^{33}$ is equal to $1-Q^{11}-Q^{22}.$

In the case (\ref{a2}) we have to consider  integrals of motion whose commutators with $P_1$ and $P_2$ are reduced to  $Q^{\mu\nu}$ with $\mu, \nu<3$ and  $Q^{33}$. They are listed in the following formula:
 \begin{gather}\la{a5} Q^{a}=\{P_a,D\}+\eta^a,\ \ \tilde Q^a=\{P_3,L_a\}+\tilde \eta^a.\end{gather}

 Thus to find the PDM systems which are invariant with respect to algebra e(2) and admit second order integrals of motion it is sufficient to fix such of them which admit the integrals (\ref{a3}) and  (\ref{a5}). Such systems can admit some additional symmetries whose calculation for {\it known} system is a rather simple problem.

 We realize the presented algorithm and find three systems with rather extended sets integrals of motion which are presented in Table 5.  The calculations requested   of solving the determining equations corresponding to  (\ref{a3}) and  (\ref{a5}) are rather straightforward. In particular, for the case of operator $Q^{31}$ (\ref{a3}) equations (\ref{mmmu1}) and (\ref{mmmu7}) are reduced to the following systems:
 \begin{gather*}f_3=2fg_1,\ \ fg_2=0,\  \ gf_3=fg_3\end{gather*}
 and
 \begin{gather*}V_3=f\eta_1^{31},\ \ f\eta_2^{31}=0,\  \ gf_3=f\eta_3^{31}\end{gather*}
 correspondingly. These systems are solved by the following functions:
 \begin{gather}\la{a6}f=\frac{c_1}{x_3+c_2}, \ \ g=\frac{x_1+c_3}{x_3+c_2},\ \ V=\frac{c_4}{x_3+c_2}+c_5,\ \ \eta=2(2+c_3)V+c_6\end{gather}
 where $c_1, c_2, ...,c_6$ are integration constants. Scaling and shifting the spatial variables  we can reduce $c_1$ to the unity and $c_2, c_3$ to zero. Constants $c_5$ and $c_6$ also can be nullified since they are added to the functions which are defined up to constant shifts. As a result we come to the system presented in Item 1 of Table 5. However, it is necessary to search for the additional integrals of motion admitted by this system, i.e., to solve the determining equations including the generic Killing tensor and functions $f$ and $V$ fixed in (\ref{a6}).

%\newpage
\vspace{3mm}

Table 5. Inverse masses and potentials for  equations admitting symmetry algebra $e(2)$

\vspace{4mm}

\begin{tabular}{c c c c}
\hline
No&$f$&$V$&\text{Integrals of motion}\\
\hline\\

1.&\vspace{2mm}$\frac1{x_3}$&$\frac{c}{x_3}$&$\begin{array}{c}2P_3P_2-(x_2\cdot H)-c\frac{x_2}{x_3} ,\ \
2P_3P_1-(x_1\cdot H)-c\frac{x_1}{x_3}
 ,\\\{P_3,D\}-\frac12((\tilde r^2+4x_3^2)\cdot H)-\frac{c(\tilde r^2-4x_3^2)}{2x3},\\
\{P_1,L_2\}+\{P_2,L_1\}+\frac12((x_1^2- x_2^2)\cdot H)+\frac{c(x_1^2-x_2^2)}{2x_3},
\\\{P_1,L_1\}-\{P_2,L_2\}+(x_1x_2\cdot H)+\frac{c x_1x_2}{x_3}

\end{array}$\\

2.&\vspace{2mm}$\frac{x_3^2}{x_3^2-1}$&$\frac{cx_3^2}{x_3^2-1}$&$\begin{array}{c}
\{P_3,L_1\}+(\frac{x_2}{x_3^2}\cdot H)+\frac{cx_2}{x_3^2-1},\\\{P_3,L_2\}-(\frac{x_1}{x_3^2}\cdot H)-\frac{cx_1}{x_3^2-1},

\\\{P_1,D\}-(x_1\cdot H)+\frac{cx_1(x_3^2-2)}{x_3^2-1},

\\ \{P_2,D\}-(x_2\cdot H)+\frac{cx_2(x_3^2-2)}{x_3^2-1},

\\ \{P_3,K_3\}+(\frac{x_3^4+\tilde r^2}{x_3^2}\cdot H)+\frac{\tilde r^2+2-x_3^4}{x_3^2-1},\\

\{P_1,K_1\}+(x_1^2\cdot H)-\frac{c\tilde x_1^2(x_3^2-2)}{x_3^2-1},\\

\{P_2,K_2\}+(x_2^2\cdot H)-\frac{c\tilde x_2^2(x_3^2-2)}{x_3^2-1},\\

\{P_1,K_2\}+\{P_2,K_1\}+2(x_1x_2\cdot H)-\frac{2cx_1x_2(x_3^2-2)}{x_3^2-1} \end{array}$\\

3.&\vspace{2mm}$\frac{x_3^2}{x_3^2+1}$&$\frac{cx_3^2}{x_3^2+1}$&$\begin{array}{c}
\{P_3,L_1\}-(\frac{x_2}{x_3^2}\cdot H)-\frac{cx_2}{x_3^2},

\\\{P_3,L_2\}+(\frac{x_1}{x_3^2}\cdot H)+\frac{cx_1}{x_3^2},

\\\{P_1,D\}-(x_1\cdot H)+\frac{cx_1(x_3^2+2)}{x_3^2+1},

\\ \{P_2,D\}-(x_2\cdot H)+\frac{cx_2(x_3^2+2)}{x_3^2+1},\\

\{P_3,K_3\}+(\frac{x_3^4-\tilde r^2}{x_3^2}\cdot H)+\frac{2-\tilde r^2-x_3^4}{x_3^2+1},\\

\{P_2,K_2\}+(x_2^2\cdot H)-\frac{c\tilde x_2^2(x_3^2+2)}{x_3^2+1},\\\{P_1,K_1\}+(x_1^2\cdot H)-\frac{c x_1^2(x_3^2+2)}{x_3^2+1},\\ \{P_1,K_2\}+\{P_2,K_1\}+2(x_1x_2\cdot H)-\frac{2cx_1x_2(x_3^2+2)}{2(x_3^2+1)} \end{array}$\\

4\vspace{1.5mm}&$x_3^2$&$c$&$P_1,\  P_2, \ K_1,  \ K_2, D, L_3$\\

\hline \hline
\end{tabular}

\vspace{3mm}
\section{Discussion}

We classify inequivalent quantum mechanical systems with position dependent masses which admit
second order integrals of motion and  three parametric symmetry groups. The classification results are summarized in Tables 3, 4 and 5. In addition, we present the superintegrable systems which are supposed to admit at least one Lie symmetry, namely, the symmetry with respect to scaling of the dependent and independent variables, see Tables 1 and 2.

As it was indicated in \cite{NZ}  there are three inequivalent three parametric Lie groups which can be admitted by the PDM Schr\"odinger equation, namely, the rotation group SO(3), the Lorentz group in (1+2)-dimensional space SO(1,2) and the Euclid group in 2d space  E(2).

We believe that the PDM systems  invariant with respect to groups SO(1,2) and E(2) are classified in the present paper for the first time.

Superintegrable PDM systems with the rotational symmetries   have been discussed in numerous papers, see \cite{Bala2}, \cite{Rag1}, \cite{Ra}, \cite{Ca} and references cited therein. A formal complete classification of such quantum mechanical systems admitting second order integrals of motion was presented in \cite{154}. In the present paper we revise the results of this classification and present its results
 in a compact form and in the only table, namely, Table 4 whenever in
 \cite{154} you can find two rather extended tables which, however,  include a lot of useful information concerning the
 supersymmetry and integrability of the discussed systems.

  Notice that the systems presented in the same item of  Table 4 and differ only by the value of
arbitrary parameters  in fact are essentially different. In particular they can possess different supersymmetry \cite{154}.

To solve the classification problems we use a specific representation of the Hamiltonians and integrals of motion fixed in equations (\ref{H}) and (\ref{Q}). Being mathematically equivalent to other representations with another orders of differentials and functions (compare (\ref{A1}) and (\ref{H})) they led to maximally compact and simple systems of the determining equations for the arbitrary elements $V$ and $f$.

The next natural steps are to classify superintegrable systems admitting two-parametric symmetry
groups and at least a one one-parametric symmetry group. Just such  systems but in two dimensions are
used and studied in numerous papers, see, e.g., \cite{kama, kaka}.

Notice that the present paper includes some important elements of such generalized analysis. Indeed,
in Tables 1 and 2  the results of the  classification of superintegrable systems invariant with
respect to the one parametric group of dilatation transformations is presented. Among them are rather
exotic systems whose potentials are arbitrary order polynomials in $\frac{x_1}{x_3}$ and
$\frac{x_2}{x_3}$ presented in Item 8 of Table 1. However, this classification is restricted to the integrals of motion which, up to scalar terms   belong to the enveloping algebra of algebra c(3).

The total number of the inequivalent one- and two- parametric Lie  groups which can be  admitted by
quantum mechanical PDM systems is not too large. In accordance with the results of paper \cite{NZ}
there exist five two parametric and five one parametric groups which can be accepted by the 3d
quantum mechanical systems with PDM. Among them there are  the groups generated by the following
pairs of infinitesimal operators belonging to the list presented in (\ref{QQ}):
\begin{gather}\la{101}<D, P_3>,\quad <D, L_3>, \quad <P_1,P_2> .\end{gather}

The superintegrable systems invariant with respect to these groups are partially classified in the present
paper. Indeed, the systems admitting the algebras spanned on $<D, P_3>$  and $<P_1,P_2>$ are
presented in Tables 3 and 5. Moreover Tables 1 and 2 include the systems admitting
the algebra $<D, L_3>$, see Item 6 of Table 1 and Items 1, 7, 8, 11, 12 of Table 2. In other words, we
present an essential  part of superintegrable systems admitting two parametric symmetry groups and the
systems admitting one out of five possible one parametric groups. We plane to complete this
classification in the following paper.

{\bf Acknowledgement} I am indebted with Universit\'a del Piemonte Orientale and Dipartimento di Scienze e Innovazione Tecnologica for the extended stay as Visiting Professor.

\end{document}